\documentclass[twoside,twocolumn,9pt]{article}
\usepackage{extsizes}
\usepackage[super,sort&compress,comma]{natbib} 
\usepackage[version=3]{mhchem}
\usepackage[left=1.5cm, right=1.5cm, top=1.785cm, bottom=2.0cm]{geometry}
\usepackage{balance}
\usepackage{mathptmx}
\usepackage{sectsty}
\usepackage{graphicx} 
\usepackage{lastpage}
\usepackage[format=plain,justification=justified,singlelinecheck=false,font={stretch=1.125,small,sf},labelfont=bf,labelsep=space]{caption}
\usepackage{float}
\usepackage{fancyhdr}
\usepackage{fnpos}
\usepackage[english]{babel}
\addto{\captionsenglish}{%
  
}
\usepackage{array}
\usepackage{droidsans}
\usepackage{charter}
\usepackage[T1]{fontenc}
\usepackage[usenames,dvipsnames]{xcolor}
\usepackage{setspace}
\usepackage[compact]{titlesec}
\usepackage{hyperref}

\usepackage{epstopdf}

\definecolor{cream}{RGB}{222,217,201}

\begin{document}

\pagestyle{fancy}
\thispagestyle{plain}
\fancypagestyle{plain}{
\renewcommand{\headrulewidth}{0pt}
}

\makeFNbottom
\makeatletter
\renewcommand\LARGE{\@setfontsize\LARGE{15pt}{17}}
\renewcommand\Large{\@setfontsize\Large{12pt}{14}}
\renewcommand\large{\@setfontsize\large{10pt}{12}}
\renewcommand\footnotesize{\@setfontsize\footnotesize{7pt}{10}}
\makeatother

\renewcommand{\thefootnote}{\fnsymbol{footnote}}
\renewcommand\footnoterule{\vspace*{1pt}%
\color{cream}\hrule width 3.5in height 0.4pt \color{black}\vspace*{5pt}} 
\setcounter{secnumdepth}{5}

\makeatletter 
\renewcommand\@biblabel[1]{#1}            
\renewcommand\@makefntext[1]%
{\noindent\makebox[0pt][r]{\@thefnmark\,}#1}
\makeatother 
\renewcommand{\figurename}{\small{Fig.}~}
\sectionfont{\sffamily\Large}
\subsectionfont{\normalsize}
\subsubsectionfont{\bf}
\setstretch{1.125} 
\setlength{\skip\footins}{0.8cm}
\setlength{\footnotesep}{0.25cm}
\setlength{\jot}{10pt}
\titlespacing*{\section}{0pt}{4pt}{4pt}
\titlespacing*{\subsection}{0pt}{15pt}{1pt}

\fancyfoot{}
\fancyfoot[LO,RE]{\vspace{-7.1pt}\includegraphics[height=9pt]{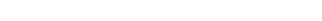}}
\fancyfoot[CO]{\vspace{-7.1pt}\hspace{11.9cm}\includegraphics{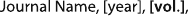}}
\fancyfoot[CE]{\vspace{-7.2pt}\hspace{-13.2cm}\includegraphics{head_foot/RF}}
\fancyfoot[RO]{\footnotesize{\sffamily{1--\pageref{LastPage} ~\textbar  \hspace{2pt}\thepage}}}
\fancyfoot[LE]{\footnotesize{\sffamily{\thepage~\textbar\hspace{4.65cm} 1--\pageref{LastPage}}}}
\fancyhead{}
\renewcommand{\headrulewidth}{0pt} 
\renewcommand{\footrulewidth}{0pt}
\setlength{\arrayrulewidth}{1pt}
\setlength{\columnsep}{6.5mm}
\setlength\bibsep{1pt}

\makeatletter 
\newlength{\figrulesep} 
\setlength{\figrulesep}{0.5\textfloatsep} 

\newcommand{\topfigrule}{\vspace*{-1pt}%
\noindent{\color{cream}\rule[-\figrulesep]{\columnwidth}{1.5pt}} }

\newcommand{\botfigrule}{\vspace*{-2pt}%
\noindent{\color{cream}\rule[\figrulesep]{\columnwidth}{1.5pt}} }

\newcommand{\dblfigrule}{\vspace*{-1pt}%
\noindent{\color{cream}\rule[-\figrulesep]{\textwidth}{1.5pt}} }

\makeatother

\twocolumn[
  \begin{@twocolumnfalse}
\vspace{1em}
\sffamily
\begin{tabular}{m{4.5cm} p{13.5cm} }

\includegraphics{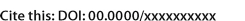} & \noindent\LARGE{\textbf{Bond breaking and making in mixed clusters of fullerene and coronene molecules following keV-ion impact }} \\
\vspace{0.3cm} & \vspace{0.3cm} \\

 & \noindent\large{Naemi Florin,$^{\ast}$\textit{$^{a}$} Alicja Domaracka,\textit{$^{b}$} Patrick Rousseau,\textit{$^{b}$} Michael Gatchell,\textit{$^{a}$} and Henning Zettergren\textit{$^{a}$}} \\

\includegraphics{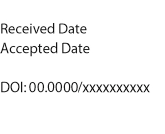} & \noindent\normalsize{We have performed classical molecular dynamics simulations of 3 keV Ar + $(\mathrm{C}_{24}\mathrm{H}_{12})_n(\mathrm{C}_{60})_{m}$ collisions where $(n,m)=(3,2), (1,4), (9,4)$ and $(2,11)$. The simulated mass spectra of covalently bound reaction products reproduce the main features of the corresponding experimental results reported by Domaracka \textit{et al., PCCP}, 2018, \textbf{20}, 15052. The present results support their conclusion that molecular growth is mainly driven by knockout where individual atoms are promptly removed in Rutherford type scattering processes. The so formed highly reactive fragments may then bind with neighboring molecules in the clusters producing a rich variety of growth products extending up to sizes containing several hundreds of atoms, and here we show examples of such structures. In addition, knocked out atoms may be absorbed such that e.g. hydrogenated coronene and fullerene molecules are formed.} 
\\
\end{tabular}

 \end{@twocolumnfalse} \vspace{0.6cm}

  ]

\renewcommand*\rmdefault{bch}\normalfont\upshape
\rmfamily
\section*{}
\vspace{-1cm}


\footnotetext{\textit{$^{a}$~Department of Physics, Stockholm University, 106 91 Stockholm, Sweden. E-mail: naemi.florin@fysik.su.se}}
\footnotetext{\textit{$^{b}$~Normandie Univ, ENSICAEN, UNICAEN, CEA, CNRS, CIMAP, 14000 Caen, France. }}

\footnotetext{\dag~Electronic Supplementary Information (ESI) available: Additional figures for small clusters containing five molecules. See DOI: 10.1039/cXCP00000x/}







\section{Introduction}

The formation and breakage of bonds are at the heart of chemical reactions and govern the evolution of molecules in terrestrial and extraterrestrial environments.\cite{Gatchell:2022aa}  In recent years, there have been numerous studies reporting molecular growth processes in gas-phase molecular clusters ignited by different ionization and excitation agents (photon, electron, and ion impact). Examples include peptide bond formation in clusters of amino acids,\cite{Rousseau:2020aa,Licht:2023aa} formations of aromatic rings in clusters of small carbon chains,\cite{Gatchell:2017aa} and growth of complex carbon nanostructures in pure and mixed clusters of Polycyclic Aromatic Hydrocarbons (PAHs) and fullerenes.\cite{Zettergren:2013vp,Seitz:2013wc,Delaunay:2015aa,Delaunay:2018aa,Gatchell:2014aa,Domaracka:2018} The cluster environment is key here as it allows the excitation energy to be shared among its molecular constituents such that the reaction products are formed sufficiently cold to survive on extended timescales (microseconds and beyond). The experiments reported so far have used various mass spectrometry techniques to provide information on the mass-to-charge ratios of the reaction products, which sometimes, as in the case of fullerenes, reveal the preferred cluster packing types.\cite{Martin:1993aa,Branz:2002ab} In general, however, they do not provide any structural information of the cluster structures before they are exposed to radiation in different forms, nor for any new species formed following such interactions. Complementary theoretical studies aiming to determine inherent cluster properties such as their stabilities and structures are thus needed to characterize the experimental targets. Pioneering theoretical studies have for instance shown that the most stable PAH clusters contain single or multiples stacks depending on the cluster size\cite{Rapacioli:2005:2}, while cluster of fullerenes prefer icosahedral packing. This is consistent with mass spectrometry observations of so-called magic numbers corresponding to particularly stable cluster sizes.\cite{Doye:1996aa,Doye:2001aa}  These structures may then be used as input in studies of intra-cluster bond breaking and formation processes by means of, e.g., classical molecular dynamics simulations to reveal detailed information about the reaction pathways and final reaction products (structures and internal energies). 

  \label{fgr:example}

\begin{figure}[h]
\centering
  \includegraphics[width=0.49\columnwidth]{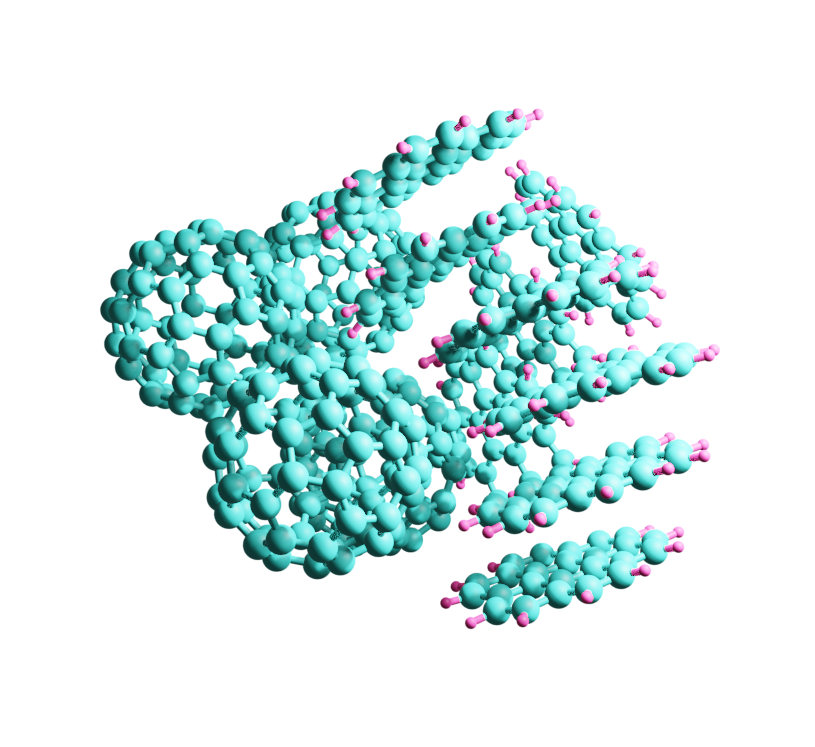}
  \includegraphics[width=0.49\columnwidth]{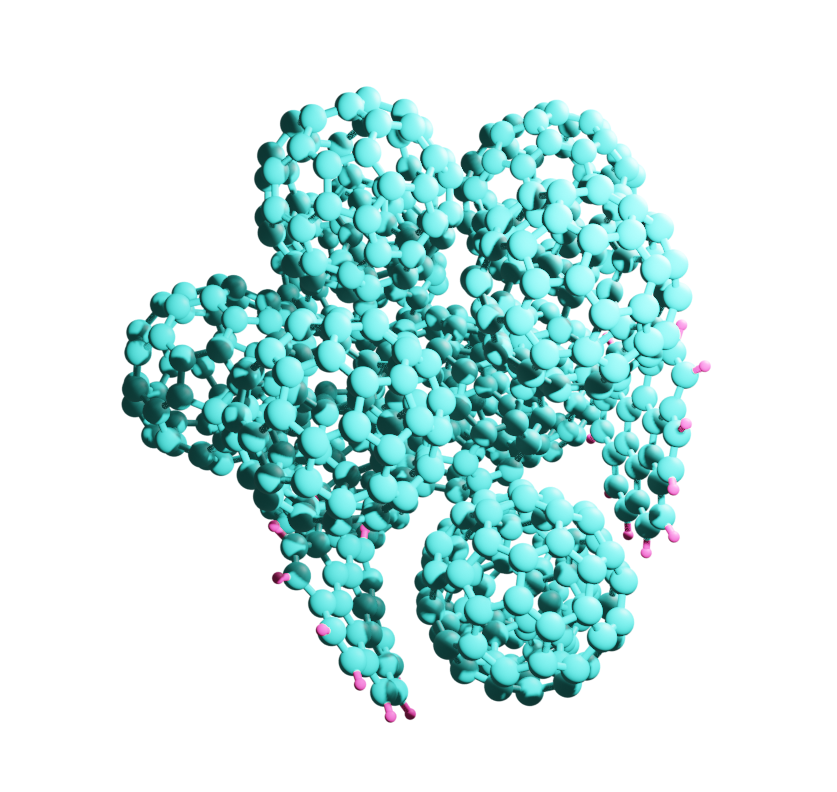}
  \caption{The most stable (C$_{24}$H$_{12}$)$_9$(C$_{60}$)$_4$ and (C$_{24}$H$_{12}$)$_2$(C$_{60}$)$_{11}$ cluster structures according to the calculations reported in Ref. \citenum{Florin:2023aa}.}
  \label{fgr:large}
\end{figure}

In this work, we have performed such simulations of collision induced chemical reactions inside of mixed clusters of PAH (C$_{24}$H$_{12}$) and fullerene (C$_{60}$) molecules. Here, we use the optimized structures from Ref. \citenum{Florin:2023aa}, see Fig.\ \ref{fgr:large} for two examples, and bombard them with 3 keV Ar-atoms mimicking the conditions in the experimental study by Domaracka \emph{et al}.\cite{Domaracka:2018} There, they showed that a large variety of new covalently bound reaction products are formed in the collisions, and that their yields depend on the mixing of C$_{24}$H$_{12}$ and C$_{60}$ in the cluster targets. For clusters containing low amounts of C$_{60}$, the molecular growth products are mainly coronene-based and there are no signs of reactions products containing C$_{60}$-molecules alone, while growth products containing both species become important as the C$_{60}$ content in the clusters is increased. The newly formed products were attributed to so-called knockout-driven reactions as have been observed to drive the ion-induced chemistry in, e.g., pure clusters of PAHs and fullerenes.\cite{Zettergren:2013vp,Seitz:2013wc,Delaunay:2015aa,Delaunay:2018aa} Single or multiple atoms are then promptly removed in Rutherford-type scattering processes forming highly reactive fragments inside the clusters that readily form new bonds with neighboring molecules.\cite{Gatchell:2016aa} These types of molecular growth processes are particularly important in the so-called nuclear stopping regime, i.e.\ for heavy keV-ion impact (velocities \emph{below} a few hundred km/s). In contrast, lighter and faster keV projectiles (velocities \emph{above} a few hundred km/s) mainly deposit energy to the molecular electron clouds (electron stopping) that typically lead to strong cluster heating followed by evaporation of intact molecules, albeit with a small probability for knockout driven reactions.\cite{Johansson:2011ab,Zettergren:2013vp,Gatchell:2014aa,Delaunay:2015aa} In case the molecular building blocks are not as rigid as fullerenes and PAHs, as in the case of, e.g., small carbon chains and amino acids, molecular growth processes are more likely to also occur following distant electron transfer processes. Ionization is often the main driving force under such circumstances and the results are then similar to those observed in electron impact induced molecular growth. 

Here, we show that the results by Domaracka \emph{et al.}\cite{Domaracka:2018}\ can be rationalized by knockout-driven reactions, which we demonstrate through the close agreement between their experimental and our simulated mass spectra for cluster targets with different PAH-fullerene mixing ratios and cluster sizes. The present results provide information on the structures of the covalently bound reaction products, which range in complexity from hydrogenated coronene and fullerene molecules to complex hydrocarbon and pure carbon structures containing hundreds of atoms.

\section{Methods}

The MD simulations were carried out using the open source MD software LAMMPS (Large-scale Atomic/Molecular Massively Parallel Simulator).\cite{Plimpton:1995aa,lammps_web} Our approach follows that previously used to successfully model collisions between keV atoms and pure PAH or fullerene clusters\cite{Delaunay:2015aa,Delaunay:2018aa} and is briefly described here. The AIREBO (Adaptive Intermolecular Reactive Empirical Bond Order) potential\cite{Stuart:2000aa,Brenner:2002aa} was used to describe bonding interactions between atoms within each PAH and fullerene molecules, as well as to describe the long-range interactions between molecules in the clusters. This type of reactive force field allows bonds to be broken and formed dynamically in a realistic manner during a simulation run, which is essential for this type of study. The interactions between the projectile and the target were governed by the ZBL (Ziegler-Biersack-Littmark) potential.\cite{zbl_pot_book} The ZBL potential describes nuclear scattering processes by treating two colliding atoms as point charges screened by their respective electron clouds.

As targets, we used optimized geometries identified in our earlier study on mixed coronene-C$_{60}$ clusters.\cite{Florin:2023aa} From those results, we selected the lowest energy structures for four different cluster compositions, $(\mathrm{C}_{24}\mathrm{H}_{12})_n(\mathrm{C}_{60})_{m}$ where ($n,m$)=(3,2), (1,4), (9,4), and (2,11) as targets for our simulations. These were selected as they represent two distinct cluster sizes---5 and 13 molecules in total, respectively---and four different mixing ratios that are close to those reported by Domaracka \emph{et al.}\cite{Domaracka:2018}

In each simulation, a whole cluster as a unit was centered and randomly orientated at the origin of a cubic box with 200\,{\AA} sides. We used 3\,keV argon atoms as projectiles, which were fired from a starting point at a fixed $z=15$\,{\AA} distance from the target center and parallel to the $z$ axis. The initial $x$ and $y$ coordinates of the projectile was randomized in each run within a square that was slightly larger than the projected size of the target cluster. This setup ensured that all orientations and impact parameters within the chosen square cross section were sampled with correct probabilities. To improve computational efficiency, runs where the projectile would pass outside the cluster and transfer only a negligible amount of energy (below about 1\,eV) were discarded before they were carried out. The simulations that were run used a time step of $5\times 10^{-17}$\,s for $2\times 10^5$ steps, giving a total simulation time of 10\,ps ($10^{-11}$\,s).

Due to the force fields used, the simulations are limited to only describing nuclear scattering processes between neutral projectiles and targets. For the projectile used here, 3\,keV Ar, electronic stopping will be significantly weaker than nuclear stopping\cite{zbl_pot_book} and its influence on the reactivity of fragments within the clusters can safely be neglected.\cite{Delaunay:2015aa} As for the charge, coronene and C$_{60}$ molecules are large, stable molecules with many delocalized electrons. This gives them properties (e.g., dissociation and binding energies) that change very little between neutral systems and ions with low charge states. Our results are thus directly comparable to the experimental results for cationic species. This has also been verified by past comparisons between experiment and theory.\cite{Delaunay:2015aa,Delaunay:2018aa,Wang:2014aa}

\begin{figure*}[htb]
\centering
  \includegraphics[width=\textwidth]{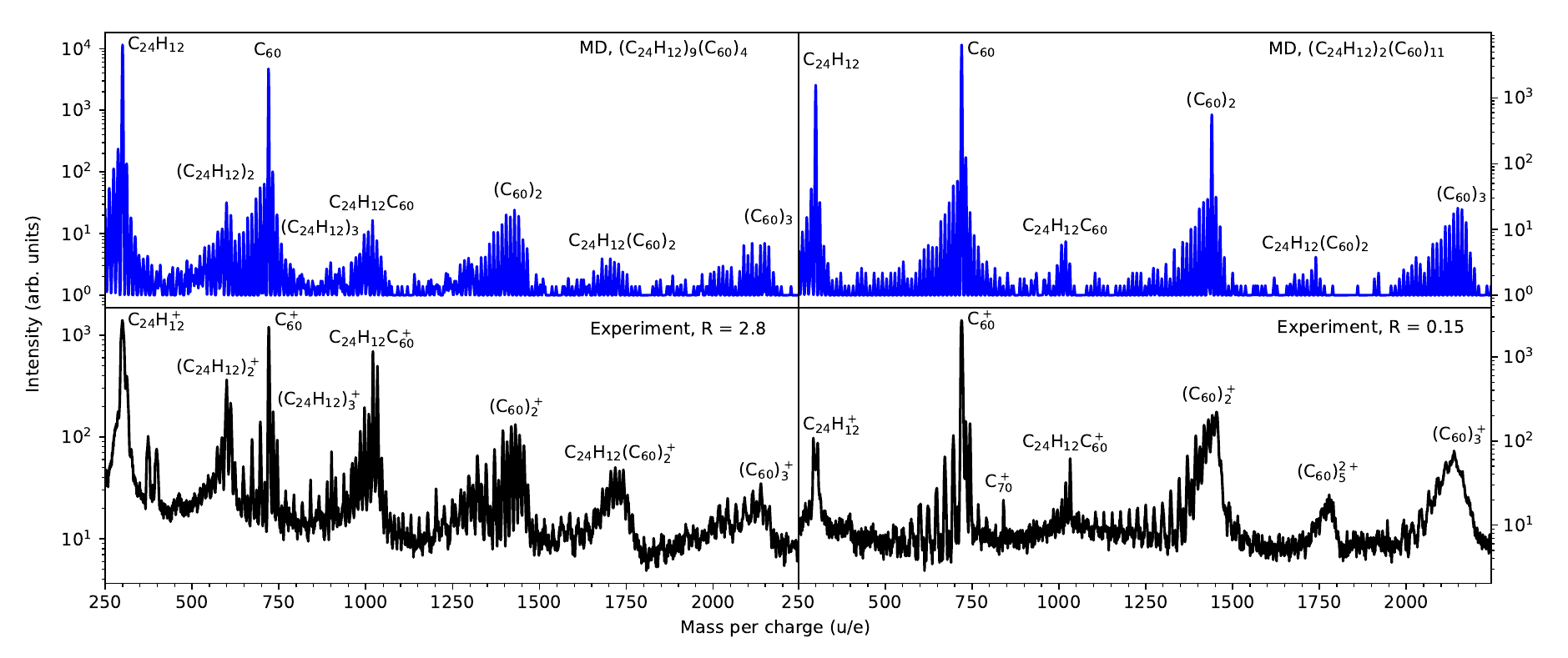}
  \caption{Upper panels: Mass spectra from molecular dynamics simulations of 3 keV Ar-projectiles colliding with $(\mathrm{C}_{24}\mathrm{H}_{12})_9(\mathrm{C}_{60})_{4}$ (left) and $(\mathrm{C}_{24}\mathrm{H}_{12})_2(\mathrm{C}_{60})_{11}$ (right). Lower panels: Experimental mass-to-charge spectra for 3 keV Ar$^+$ colliding with broad distributions of mixed clusters of coronene and fullerene molecules \citenum{Domaracka:2018}. The coronene/fullerene mixing ratios are R=2.8 and R=0.15 in the left and right panel, respectively.}
  \label{fgr:masses}
\end{figure*}

\section{Results and discussion}


We find that collisions with the larger cluster targets ($N=13$ molecules) give reaction products extending up to maximum sizes that are slightly larger than in collisions with the smaller clusters ($N=5$), which is expected since more hydrocarbon material is available for bond forming reactions in larger clusters. Apart from that, similar type of reaction products are produced in collisions with all four cluster targets, suggesting that the growth mechanisms are the same. The mixing ratio and cluster size instead merely affects the branching ratios for the final distributions of reaction products. In the following we will primarily focus on the growth mechanisms and the structures of the products and we therefore only show the results for clusters containing 13 molecules. The corresponding results for the smaller clusters are given in the supporting information.

\subsection{Simulated and experimental mass spectra}

Figure \ref{fgr:masses} shows the simulated mass spectra for the cluster targets, $(\mathrm{C}_{24}\mathrm{H}_{12})_9(\mathrm{C}_{60})_{4}$ (upper left panel), $(\mathrm{C}_{24}\mathrm{H}_{12})_2(\mathrm{C}_{60})_{11}$ (upper right panel), together with experimental results from Domaracka \emph{et al.}\cite{Domaracka:2018}\ for collisions with mixed clusters containing mostly coronene (lower left panel) and fullerene molecules (lower right panel). In the simulated spectra, only the covalently bound reaction products are shown, while in the experiments it is not possible to distinguish those from any weakly bound cluster precursors and remnants or larger clusters having the same mass. For this reason the relative intensities of peaks at masses corresponding to intact clusters are not directly comparable between experiment and theory. The four panels display similar features with a wide range of reaction products spanning in size from the intact coronene monomer ($m=300$\,u) to slightly above that of the intact fullerene trimer ($m=2160$\,u). The most feature-rich spectrum from the simulations is for $(\mathrm{C}_{24}\mathrm{H}_{12})_9(\mathrm{C}_{60})_{4}$, which is consistent with the experimental results showing a propensity for intracluster reactions in clusters containing a higher fraction of coronene molecules (lower left panel). 

\subsection{Covalently bound intact and defected mixed dimers}

\begin{figure}[h]
\centering
  \includegraphics[width=\columnwidth]{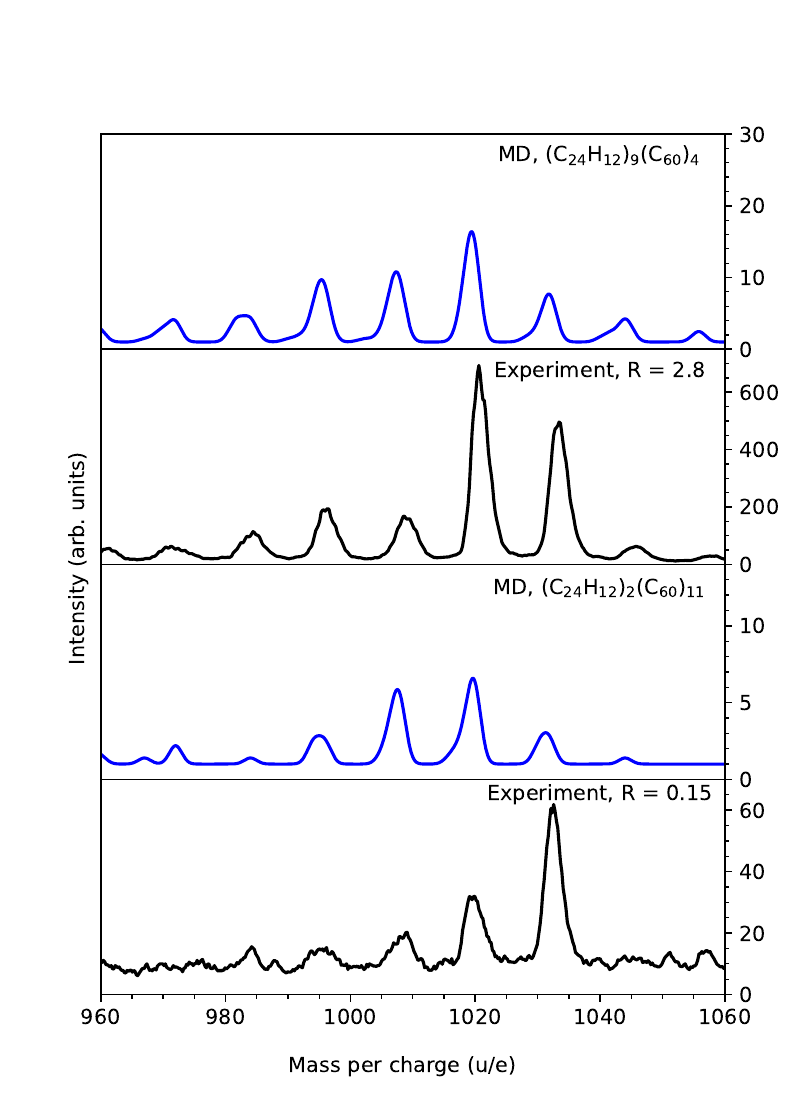}
  \caption{Comparisons of simulated (blue) and experimental\cite{Domaracka:2018} (black) mass spectra in the region of the mixed dimer. The upper and lower panels show the comparisons for clusters with high and low coronene contents, respectively. Examples of reaction products from the simulations are shown in Fig. \ref{fgr:threetypes}.}
  \label{fgr:zoomin}
\end{figure}

\begin{figure}[h]
\centering
  \includegraphics[width=\columnwidth]{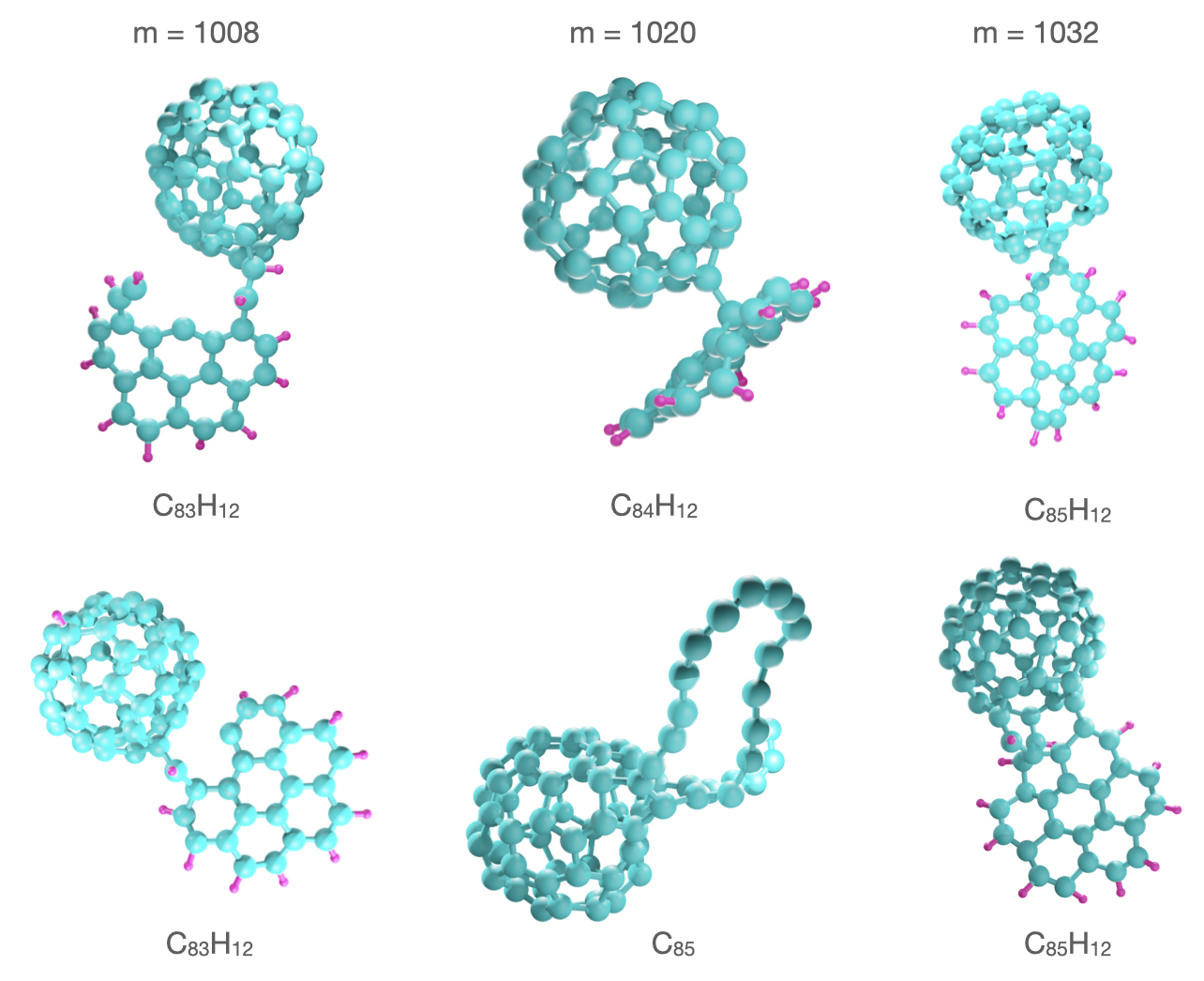}
  \caption{Examples of covalently bound reaction products with masses m=1008, 1020, and 1032.}
  \label{fgr:threetypes}
\end{figure}

In Fig.\ \ref{fgr:zoomin} we show the simulated and measured mass spectra in the region of the mixed dimer, $(\mathrm{C}_{24}\mathrm{H}_{12})(\mathrm{C}_{60})$, for which $m=1020$. 
The upper two panels show the theoretical and experimental results for targets with a high coronene content, while the lower two panels show the equivalent data for clusters with a low coronene content.
The simulated mass spectra have been convoluted with a gaussian of width $\text{FWHM}=0.5$\,u such that the peak widths are similar to those in the corresponding experimental mass spectra in order to aid the comparisons. The effective widths of the peaks in Fig.\ \ref{fgr:zoomin} are the result of the underlying distributions of the number of H atoms for fragments of similar masses. The simulations reproduce the main features of the experimental results remarkably well. For collisions with clusters containing a dominant fraction of coronene molecules the simulated and experimental mass spectra contain clear peaks corresponding to reaction products ranging in size from five carbon atoms less than the intact dimer ($m=960$\,u) up to three additional carbon atoms ($m=1056$\,u). These peaks are separated by one carbon mass, which is a clear fingerprint for knockout driven reactions with PAHs and fullerenes.\cite{Zettergren:2013vp,Delaunay:2015aa,Gatchell:2016aa} Here, highly reactive fragments from such processes form bonds with intact or defective neighbors leading to reaction products with $m<1020$\,u. Alternatively, knocked-out carbon atoms are absorbed by molecules which then react with neighbors such that the reaction products are heavier than the intact dimer ($m>1020$\,u). In addition, covalently bound reaction products with $m=1020$\,u are formed in, e.g., fusion of two intact molecules or in reactions with damaged molecules and molecules that have absorbed one or several knocked out atoms. Typical examples of these three types of reaction products are shown in Fig.\ \ref{fgr:threetypes}. The left column shows two species with mass $m=1008$\,u. The upper structure is a coronene fragment after single carbon knockout forming a single bond with an intact fullerene, while the lower is a coronene that has lost one carbon atom and one hydrogen atom and is forming a covalent bond with a fullerene that has absorbed a knocked out hydrogen atom. The middle column shows structures corresponding to the mass of an intact mixed dimer. In most cases such species are formed in reactions between damaged but non-fragmented coronenes and intact fullerenes as illustrated in the upper figure, but there are also examples where a knockout fragment reacts with a molecule that has absorbed atom(s) with the same mass as the knocked out atom(s) from the other molecule (not shown). Furthermore, species consisting purely of carbon atoms (C$_{85}$) are formed in collisions that have lead to severe damage of a fullerene that binds to a neighboring fullerene (see the lower middle panel of Fig.\ \ref{fgr:threetypes}). Finally, the right column of Fig.\ \ref{fgr:threetypes} shows structures with $m=1032$\,u corresponding to the addition of one carbon atom to an intact mixed dimer. An example of the most typical binding situation is shown in the upper panel where a displaced carbon is acting as a link between two intact molecules in a similar fashion as dumbbell fullerene dimers (C$_{121}$) observed when atomic carbon reacts with fullerenes in helium nanodroplets.\cite{Krasnokutski:2016aa} Another, more exotic, example is shown below this where the fullerene and coronene molecules are bridged by two new bonds, one where a H atom has been stripped from the PAH and one where an additional CH fragment has formed a bond between the pair.


\begin{figure}[h]
\centering
  \includegraphics[width=\columnwidth]{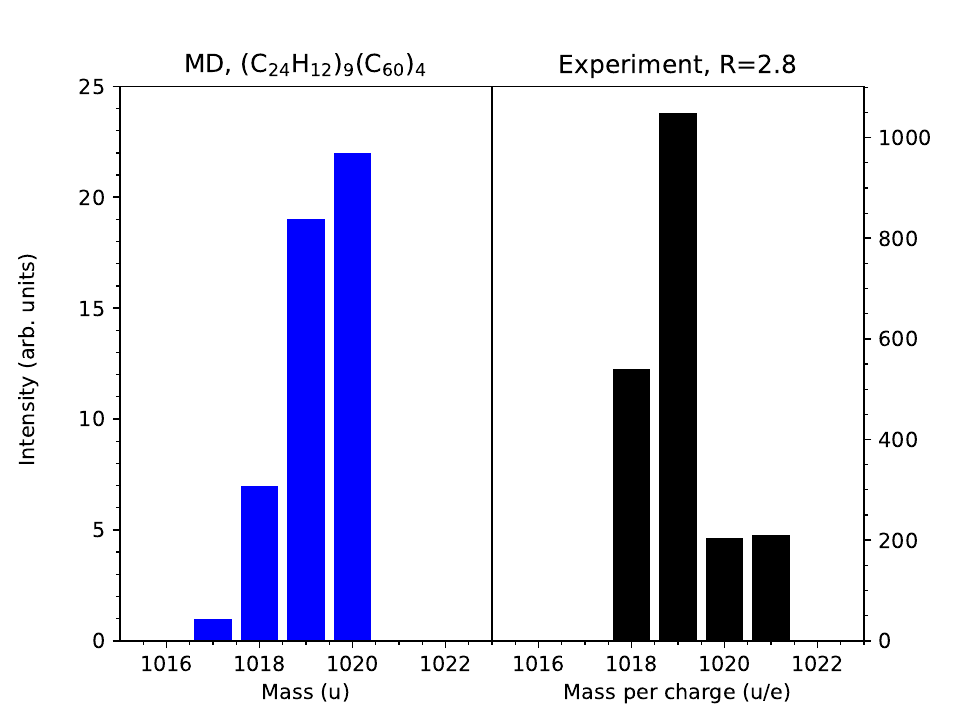}
  \caption{Relative intensities of reaction products in the vicinity of the mixed dimer (m=1020 \,u). Left: MD simulations. Right: Experimental results \citenum{Domaracka:2018}.}
  \label{fgr:zoomhydrogen}
\end{figure}

Fig.\ \ref{fgr:zoomhydrogen} shows a closer comparison of the simulated and experimental mass distribution near 1020\,u stemming from collisions with clusters containing a high fraction of coronene molecules. The experimental distribution has been extracted from the higher resolution spectrum reported in Ref.\ \citenum{Domaracka:2018} where individual peaks separated by one hydrogen mass are resolved. Peaks corresponding to one or two hydrogen atoms fewer than for an intact mixed dimer are clearly visible in the simulations and experiments. In the latter case there is also a peak at $m=1021$\,u corresponding to one additional hydrogen, which is not seen in the present simulations. We attribute this to the small cluster size in comparison with the typical ones in the experiments, i.e.\ that there is a significantly lower probability that knocked out hydrogen atoms are absorbed by a molecule in small cluster and hence secondary bond formation processes involving such species and their neighbors are extremely rare. In the simulations, there is a significantly higher fraction of $m=1020$\,u in relation to the smaller masses compared to the experiments. This suggests that the internal energies of such reaction products often are too high for them to survive on the experimental timescales. The lower-mass products in this range are expected to be more stable as hydrogen atoms have been lost from coronene, which produces reactive sites where strong bonds to neighboring molecules can be formed. The secondary loss of hydrogen after a bond-forming reaction has taken place could also assist in stabilizing reaction products on experimental timescales, an aspect not fully captured by the short timescales of our simulations.

\subsection{Dehydrogenated and hydrogenated coronene molecules}

\begin{figure}[h]
\centering
  \includegraphics[width=\columnwidth]{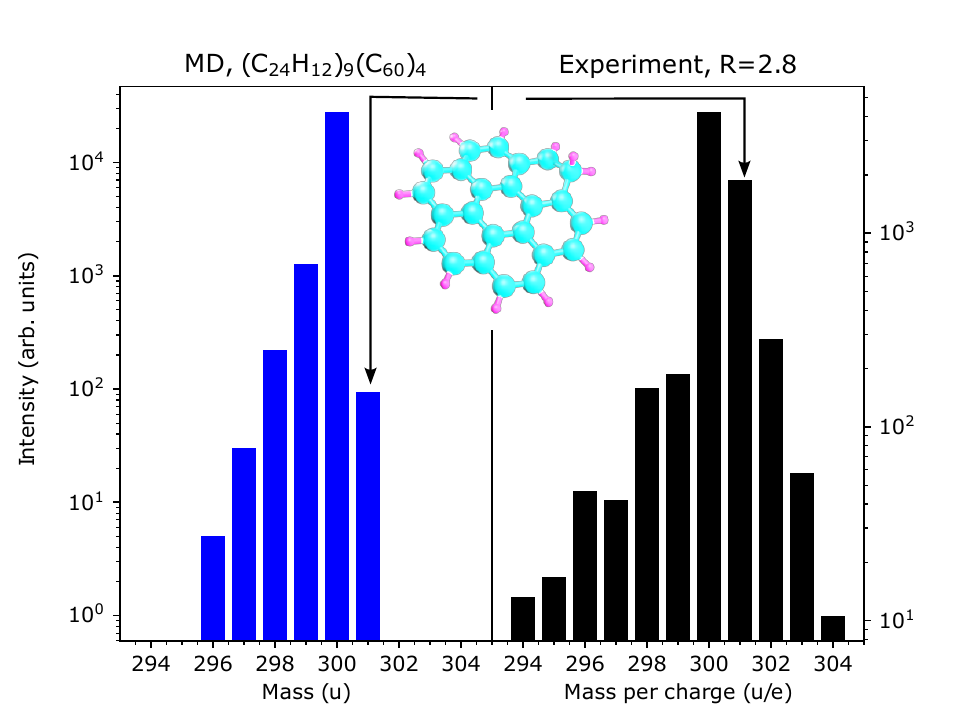}
  \caption{Relative intensities of covalently bound reaction products in the vicinity of the intact coronene molecule (m=300\,u). Left: MD simulations. Right: Experimental results \citenum{Domaracka:2018}. The inset shows the most common structure of singly hydrogenated coronene (m=301\,u) formed in the simulations.}
  \label{fgr:coronenehydrogen}
\end{figure}

\begin{figure}[h]
\centering
  \includegraphics[width=\columnwidth]{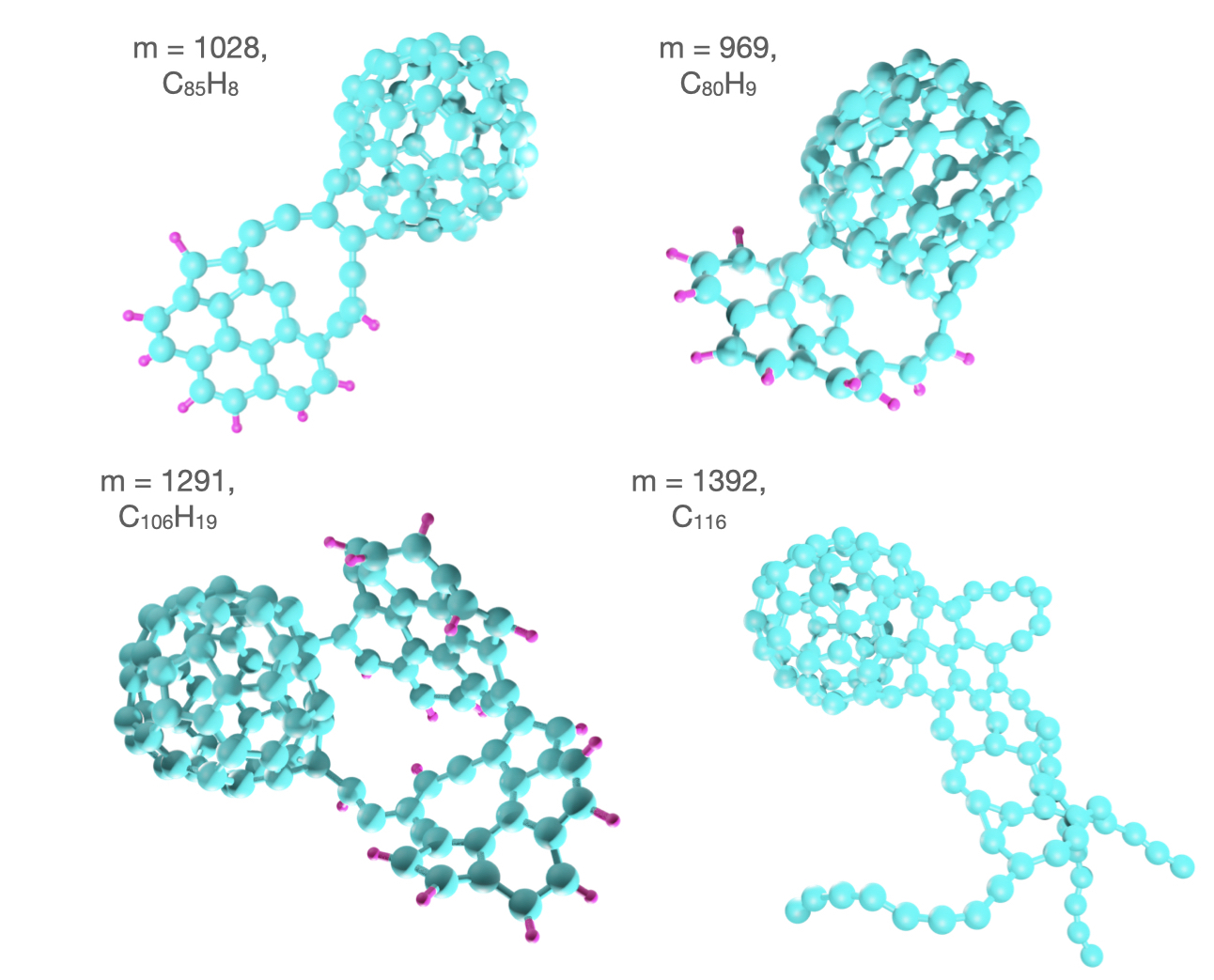}
  \includegraphics[width=5cm]{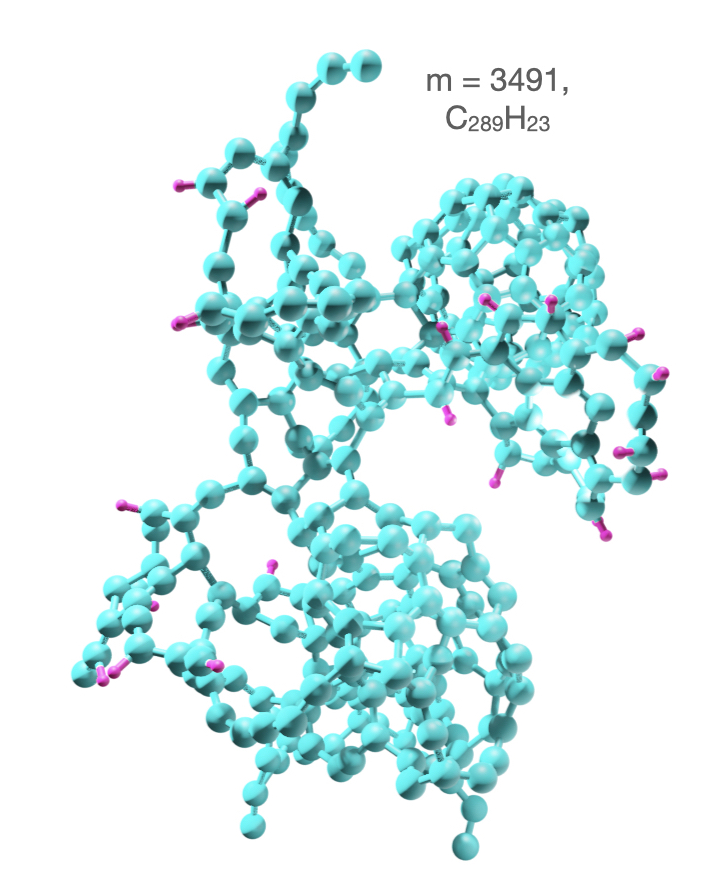}
  \includegraphics[width=5cm]{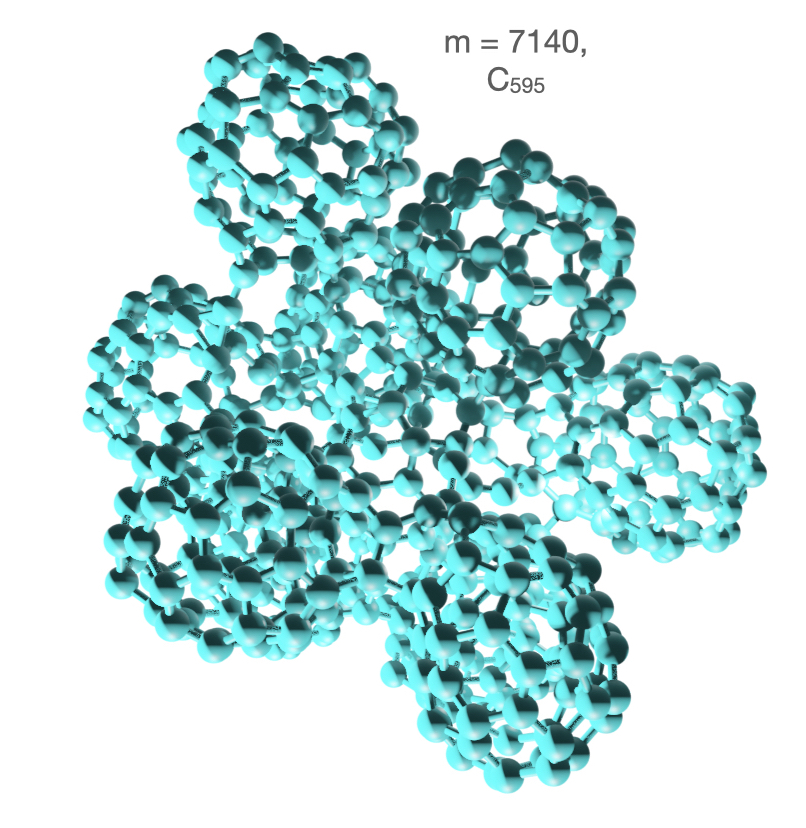}
  \caption{Examples of large covalently bound reaction products (see text).}
  \label{fgr:exotic}
\end{figure}

The conclusion that hydrogen atoms are more likely to be absorbed by their neighbors in collision with large clusters is further supported when comparing the simulated and experimental mass spectra in the region of the intact coronene molecule (see Fig.\ \ref{fgr:coronenehydrogen}). The simulated mass spectrum shows that up to four hydrogen atoms are knocked out ($m=296\ldots299$\,u) and in some cases one hydrogen atom is picked up by another coronene molecule in the cluster ($m=301$\,u). These reaction products are also seen in the experimental mass spectrum, as are products from the loss of five and six hydrogen atoms ($m=295$\,u and 296\,u) and the absorption of 2--4 additional hydrogen atoms ($m=302\ldots304$\,u) in relation to the pristine coronene molecule. The present simulations show that the hydrogen atom is predominantly attached to the outer rim of the coronene molecule such that a CH$_{2}$-group is formed (see the inset in Fig.\ \ref{fgr:coronenehydrogen}). In the experiments, more than one additional hydrogen is common and it is likely that they are also predominantly located at different positions at the outer rim. Such hydrogenated species have been suggested to contribute to formation of molecular hydrogen in the interstellar medium and may thus play key roles in star formation.\cite{Page:2009aa} In the simulations we also see that hydrogenated fullerenes are formed, which are believed to exist in circumstellar and interstellar environments.\cite{2041-8205-724-1-L39}

\subsection{Exotic reaction products}

In some cases the collisions ignite molecular growth processes where exotic structures are formed through multiple bond breaking and forming processes. These typically involve fusion of molecules that have been severely damaged by the projectile. A few examples are highlighted in Fig.\ \ref{fgr:exotic}. The upper panel shows two structures stemming from reactions involving one damaged coronene and a fullerene that to a larger extent retains its original structure. These are connected through two bonds forming large ring structures from the break-ups of two and three rings in coronene, respectively. The left structure in the second row stems from two damaged coronene molecules that are attached to a fullerene. On the right hand side there is a pure carbon cluster consisting of 116 atoms from a reaction of an intact fullerene with a large and open fullerene fragment. The bottom two panels in Fig.\ \ref{fgr:exotic} show the largest covalently bound reaction products formed in collisions with clusters with a high fraction of coronene and fullerene molecules, respectively. In the former case, C$_{289}$H$_{23}$ is produced and consists of a significant fraction (63\,\%) of the carbon atoms in the $(\mathrm{C}_{24}\mathrm{H}_{12})_9(\mathrm{C}_{60})_{4}$ precursor cluster. The fraction is even larger in collisions with  $(\mathrm{C}_{24}\mathrm{H}_{12})_2(\mathrm{C}_{60})_{11}$ where the largest reaction product, C$_{595}$ as shown in Fig.\ \ref{fgr:exotic}, contains 84\,\% of the carbon atoms in a complex that largely preserves the structure of the fullerene component of the precursor cluster. 

\section{Summary and conclusions}
We have studied molecular growth processes in weakly bound mixed clusters of coronene and fullerene molecules by means of classical molecular dynamic simulations. These are ignited by 3\,keV Ar impact and we find that:
\begin{itemize}

\item The simulated mass spectra reproduce the experimental ones reported by Domaracka \emph{et al.}\cite{Domaracka:2018} Minor differences can be attributed to i) the different timescales and hence survival probabilities probed in the simulations (up to 10 picoseconds) and in the experiments (up to microseconds), and ii) that there is a broad distribution of cluster sizes in the experiments, while there is a well-defined but rather small cluster size in the simulations.  

\item The actual reaction products do not depend strongly on the fullerene/coronene content in the clusters investigated in this study, only the branching ratios for the final states.   

\item There is a rich distribution of growth products extending in size from one mass above the intact molecular cluster building blocks to pure carbon or hydrocarbon structures containing hundreds of atoms. A plethora of new species are thus formed with different and unique spectroscopic signatures (IR-emission and absorption features). 

\item Hydrogenated coronene and fullerene molecules are readily formed. The former is believed to serve as an important reservoir for H$_2$ formation - a key ingredient for star formation.  

\item The growth products are due to knockout-driven reactions which may be important in certain astrophysical environments. Here, the keV Ar impact is analogous to the processing of material by shock waves in the vicinity of old stars\cite{Micelotta:2010ab,Micelotta:2010aa} in a similar fashion as in the study by Bernal \emph{et al.}\ where fullerenes were observed to be produced following high energy Xe-atom bombardment of a silicon carbide surface.\cite{Bernal:2019aa}
\newline
\end{itemize}
\section*{Author Contributions}

\section*{Conflicts of interest}
There are no conflicts to declare.

\section*{Acknowledgements}
 M.G. and H.Z.\ acknowledge support from the Swedish Research Council (contracts 2020-03104 and 2020-03437, respectively). This work is a part of the project ``Probing charge- and mass-transfer reactions on the atomic level'', supported by the Knut and Alice Wallenberg Foundation (Grant no.\ 2018.0028). This article is based upon work from COST Actions CA18212 -- Molecular Dynamics in the GAS phase (MD-GAS) and CA21101 -- Confined Molecular Systems: From a New Generation of Materials to the Stars (COSY), supported by COST (European Cooperation in Science and Technology). 

\bibliography{rsc.bib} 
\bibliographystyle{rsc.bst} 

\end{document}